\documentclass[a4paper,twocolumn,accepted=2024-03-29,11pt]{quantumarticle}
\pdfoutput=1

\usepackage[utf8]{inputenc}
\usepackage[english]{babel}

\usepackage[T1]{fontenc}
\usepackage{amsmath,amssymb,amsthm}
\usepackage{bbold}
\usepackage{physics}
\usepackage{multirow}
\usepackage{array}
\usepackage{xcolor}
\usepackage{dsfont}
\usepackage{txfonts}
\usepackage{cancel}
\usepackage{ulem}

\usepackage[numbers]{natbib}
\usepackage{listings}
\usepackage{hyperref}

\newtheorem{theorem}{Theorem}
\newtheorem{lemma}{Lemma}
\newtheorem{conjecture}{Conjecture}

\begin{document}

\title{Robust sparse IQP sampling in constant depth}

\author{Louis Paletta}
\affiliation{Laboratoire de Physique de l'Ecole normale supérieure, ENS-PSL, CNRS, Inria, Centre Automatique et Systèmes (CAS), Mines Paris, Université PSL, Sorbonne Université, Université Paris Cité, Paris, France}
\orcid{0009-0009-3921-0404}
\author{Anthony Leverrier}
\affiliation{Inria Paris, France}
\orcid{0000-0002-6707-1458}
\author{Alain Sarlette}
\affiliation{Laboratoire de Physique de l'Ecole normale supérieure, ENS-PSL, CNRS, Inria, Centre Automatique et Systèmes (CAS), Mines Paris, Université PSL, Sorbonne Université, Université Paris Cité, Paris, France}
\affiliation{Department of Electronics and Information Systems, Ghent University, Belgium}
\orcid{0000-0001-5909-437X}
\author{Mazyar Mirrahimi}
\affiliation{Laboratoire de Physique de l'Ecole normale supérieure, ENS-PSL, CNRS, Inria, Centre Automatique et Systèmes (CAS), Mines Paris, Université PSL, Sorbonne Université, Université Paris Cité, Paris, France}
\orcid{0000-0001-9471-6031}
\author{Christophe Vuillot}
\affiliation{Universit\'e de Lorraine, CNRS, Inria, LORIA, F-54000 Nancy, France}
\orcid{0000-0002-3445-0179}

\maketitle

\begin{abstract}
Between NISQ (noisy intermediate scale quantum) approaches without any proof of robust quantum advantage and fully fault-tolerant quantum computation, we propose a scheme to achieve a provable superpolynomial quantum advantage (under some widely accepted complexity conjectures) that is robust to noise with minimal error correction requirements. We choose a class of sampling problems with commuting gates known as sparse IQP (Instantaneous Quantum Polynomial-time) circuits and we ensure its fault-tolerant implementation by introducing the tetrahelix code. This new code is obtained by merging several tetrahedral codes (3D color codes) and has the following properties: each sparse IQP gate admits a transversal implementation, and the depth of the logical circuit can be traded for its width. Combining those, we obtain a depth-1 implementation of any sparse IQP circuit up to the preparation of encoded states. This comes at the cost of a space overhead which is only polylogarithmic in the width of the original circuit. We furthermore show that the state preparation can also be performed in constant depth with a single step of feed-forward from classical computation. Our construction thus exhibits a robust superpolynomial quantum advantage for a sampling problem implemented on a constant depth circuit with a single round of measurement and feed-forward.
\end{abstract}

\section{Introduction}

Recent progress on quantum hardware suggests that quantum processors will soon be able to outperform classical devices for some specific tasks. In the absence of fault-tolerant quantum computers, sampling problems \cite{lund2017quantum,movassagh2023hardness} appear to be a promising avenue to demonstrate such a quantum advantage since they can be solved with reasonably small circuits. In sampling problems, given some family $\mathfrak{C}$ of quantum circuits on $N$ quantum registers, the goal is to sample from the output distribution $p_C$ for any circuit $C \in \mathfrak{C}$.
Well-known examples of circuit families include linear optical circuits in the case of BosonSampling~\cite{aaronson2011computational}, random quantum circuits~\cite{bouland2019complexity} and Instantaneous Quantum Polynomial-time (IQP) circuits~\cite{shepherd2009temporally}.
The original idea behind these proposals was that quantum processors can in principle sample from the corresponding distributions, while it is widely believed that classical computers cannot complete the same task efficiently. The caveat, however, is that current quantum processors are not equipped with fault-tolerance, and will instead output noisy samples, thus only solving a noisy version of the initial sampling problem. Unfortunately, the evidence for the classical hardness of this problem is thinner, and recent works have cast some serious doubts on the possibility of demonstrating a quantum advantage with this approach~\cite{aharonov2023polynomial,zhou2020limits,napp2022efficient,gao2024limitations,fefferman2023effect}.

A potential strategy to address the issue of noise is to focus on problems for which it is possible to add some level of fault-tolerance, in an intermediate manner between Noisy Intermediate-Scale Quantum (NISQ) processors available in the near term~\cite{preskill2018quantum} and universal fault-tolerant quantum computation. We list potential approaches to such robust quantum advantage in Table~\ref{table}.
The fact that IQP circuits are a non-universal class of circuits make them a good candidate in this respect since they are easier to make fault-tolerant. In particular, they can bypass the limitations of the Eastin-Knill theorem which states that a universal gate set cannot be implemented with transversal gates~\cite{eastin2009restrictions}. 
In this work, we show how to perform a fault-tolerant version of (sparse) IQP sampling with a constant-depth quantum circuit and with a space overhead that is only polylogarithmic in the width of the original circuit. We note that \cite{mezher2020fault} addressed a similar question for a different sampling problem, but the constant depth was obtained at the price of a polynomial overhead in terms of qubits because of differences in the initial computational problem and of the magic state distillation protocol necessary to its fault-tolerant implementation. In addition, it neglects some polynomial-time classical computation necessary for the error correction but during which errors can accumulate, while we bring down the complexity of error correction to polylogarithmic-time, making it less of an issue for future implementations. We note that similar computation times, for correcting a surface code of logarithmic size for instance, are often neglected in the literature.

\begin{table*}[t]
	\small
	\raggedleft
	\begin{center}
		\begin{tabular}{b{3.6cm}b{2.4cm}b{2 cm}b{2.cm}b{2.8cm}}
			\hline
			problem & space overhead & depth & advantage & assumptions \\ \hline
			factoring \cite{gidney2021factor} & polylog & poly, adapt. & superpoly & factoring hard \\
			graph state \cite{mezher2020fault} & poly & $\order{1}$, adapt. & superpoly & PH $= \infty$ \& ACH \\
			sparse IQP [this work] & polylog & $\order{1}$, adapt. & superpoly & PH $= \infty$ \& ACH \\
			magic square \cite{bravyi2020quantum} & polylog & $\order{1}$ & quasi-log & unconditional \\
			\hline
		\end{tabular}
		\caption{Potential candidates for the demonstration of robust quantum advantage. The advantage is relative between the quantum depth and its minimal classical counterpart. Factoring displays a superpolynomial advantage provided that factoring is hard classically, but requires the full machinery of fault-tolerance. Graph state sampling and sparse IQP sampling also give a large advantage, under stronger assumptions (that the Polynomial Hierarchy does not collapse, and with an Average Case Hardness conjecture) and can be implemented with an adaptive circuit of constant depth. Finally the magic square problem leads to an unconditional advantage with a non-adaptative circuit of constant depth, but only offers a logarithmic advantage compared to classical computing.}
		\label{table}
	\end{center}
\end{table*}

\subsection{Sparse IQP}

\begin{figure}[t]
	\centering
	\includegraphics[width=0.46\textwidth]{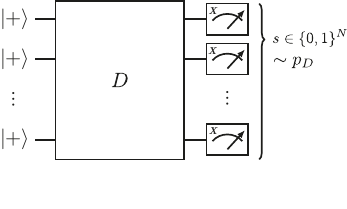}
	\caption{IQP circuits on $N$ qubits are defined by an unitary $D$ diagonal in the computational basis with state preparation and measurements performed in the Hadamard basis. For sparse IQP circuits, $D$ is a logarithmic-depth circuit consisting of $T$ and $CS$-gates.}
	\label{fig:IQP}
\end{figure}

An IQP circuit on $N$ qubits takes a very simple form (see Figure~\ref{fig:IQP}): one applies an $N$-qubit gate $D$, diagonal in the computational basis, to an initial state $|+\rangle^{\otimes N}$ and measures the resulting state in the $\left\{\ket{+},\ket{-}\right\}$ basis~\cite{bremner2011classical,bremner2016average,shepherd2009temporally}.
Here, we will focus on the \emph{sparse} variant of IQP circuits introduced in \cite{bremner2017achieving}. In this variant, the circuit $D$ is generated randomly from logarithmic-depth circuits with gate set $\{T,CS\}$: 
\begin{equation}
	T = \begin{bmatrix}
		1 & 0 \\
		0 & e^{i\pi/4} \\
	\end{bmatrix},
	\quad
	CS = \begin{bmatrix}
		1 & 0 & 0 & 0\\
		0 & 1 & 0 & 0 \\
		0 & 0 & 1 & 0 \\
		0 & 0 & 0 & i
	\end{bmatrix}.
\end{equation}
More precisely, such a circuit on $N$ qubits is generated in the following way:
\begin{itemize}
	\item a single-qubit gate $T^k$ is applied to every qubit, with $k \in \{0, \ldots, 7\}$ chosen uniformly and independently for every qubit,
	\item for every pair of qubits, a gate $CS^k$ with $k\in \{0, \ldots, 3 \}$ chosen uniformly at random, is applied with probability $\gamma \log N/N$, for some fixed parameter $\gamma>0$.
\end{itemize}
Let us denote by $\mathfrak{D}_N$ the \emph{family} of IQP circuits generated from the gate set $\{T,CS\}$. We associate to each circuit of $\mathfrak{D}_N$ its probability of being generated by the previous random process to define a \emph{distribution} over $\mathfrak{D}_N$. We call an IQP circuit picked from this distribution sparse and in the following whenever we discuss about a fraction of sparse IQP circuits we mean a fraction of circuit in the sense of the probability distribution defined above.
We note that all the considered gates commute, and can therefore be applied in any order. Given that each qubit will typically be involved in a logarithmic number of 2-qubit gates, we see that sparse IQP circuits can be implemented by circuits of average depth $\Theta(\log N)$~\cite{bremner2017achieving}.
For each sparse IQP circuit $D$, we denote by $p_D$ the probability distribution on $\{0,1\}^N$ corresponding to the output distribution of the circuit. In particular, it holds that
\begin{align}
p_D(0^N) = \sum_{z \in \{0,1\}^N} e^{i \pi/8 \left(\sum_{i<j} w_{i,j} z_i z_j + \sum_{k=1}^N v_k z_k\right) }, 
\end{align}
for some integer weights $w_{i,j}, v_k$. This quantity corresponds to an Ising model partition function, which is proven to be hard to compute in the worst case \cite{goldberg2017complexity,fujii2017commuting} and conjectured to be hard to compute on average.
We formally recall the conjecture from~\cite{bremner2017achieving}:
\begin{conjecture}[Average Case Hardness of Ising model \cite{bremner2017achieving}]\label{cj}
	Consider the partition function of the general Ising model,
	\begin{equation}
		Z(\omega) = \sum_{z\in \{\pm 1\}^N} \omega^{\sum_{i<j} w_{i,j}z_iz_j+\sum_{k=1}^N v_k z_k},
	\end{equation}
	where the exponential sum is over the complete graph on $N$ vertices, $w_{i,j}\in\mathds{R}$ and $v_k\in \mathds{R}$ are weights for edge $ij$ and vertex $k$, and $\omega \in \mathds{C}$.
	\par If the weights are chosen uniformly at random from the set $\{0, \ldots, 7\}$, then it is $\#$P-hard to approximate $\abs{Z(e^{i\pi/8})}^2$ up to multiplicative error $1/4 + o(1)$ for a $1/24$ fraction of instances, over the random choice of weights.
\end{conjecture}

The sparse IQP problem is as follows: pick a random $D \in \mathfrak{D}_N$ according to the random process described before, and output an $N$-bit string $s$ according to a distribution $q_D$ such that
\begin{align}
\|p_D - q_D\|_{\mathrm{TV}} \leq \delta,
\end{align}
where the total variation distance between two distributions $p$ and $q$ is defined as
\[ \|p-q \|_{\mathrm{TV}} := \frac{1}{2} \sum_{s \in \{0,1\}^N} |p(s) - q(s)|.\]

Assuming Conjecture \ref{cj}, and the non collapse of the Polynomial Hierarchy, a generalisation of the $P \neq NP$ conjecture widely considered to be true, Bremner \textit{et al} proved that there is no efficient classical algorithm for the sparse IQP problem. 
More precisely,
\begin{theorem}[Classical hardness of sparse IQP sampling \cite{bremner2017achieving}]\label{th:ch}
	Assuming Conjecture \ref{cj}, there exists $\delta>0$ independent of $N$ such that a constant fraction of sparse IQP circuits cannot be simulated by a polynomial-time classical algorithm up to precision $\delta$ in total variation distance unless the polynomial hierarchy collapses to its third level.
\end{theorem}

This theorem states that on average over the choice of $D$ from the probability distribution over $\mathfrak{D}_N$ defined previously, it is hard to sample classically from a distribution close to $p_D$. While a fault-tolerant quantum computer can sample efficiently from such a distribution, we do not expect that this is the case for near-term quantum processors. 
In fact, the initial proposal~\cite{bremner2017achieving} partially addressed this issue by considering a simple noise model where the quantum circuit is assumed to be ideal, except for some independent and identically distributed noise added to the classical value of the final outcomes. Unfortunately, this model is too naive and a more realistic noise model should assume that \textit{every gate} suffers from some constant level of noise. In that case, because the number of gates is of order $N \log N$ in the circuit, it is immediate that noise will accumulate through the circuit and that the level of noise per qubit cannot be assumed to be constant, independent of $N$.
Here we choose to consider a more general error model -- the \textit{local stochastic noise model}~\cite{gottesman2014faulttolerant} -- that includes well-known error models such as the independent depolarizing noise channel but also allows for local correlated errors. In this model described in Section~\ref{section:ft}, errors are applied at each gate operation and the probability that faulty locations contain a specific set $\mathcal{A}$ is upper bounded by $p^{|\mathcal{A}|}$.
In this work we propose a physical implementation of sparse IQP circuits that is robust to this kind of noise, without requiring the full machinery of fault-tolerant quantum computation.

In order to avoid multiple rounds of costly error correction, our main strategy is to make the encoded circuit of constant depth rather than logarithmic. This is challenging since the target logical circuit has logarithmic depth, and we want in addition to make it fault-tolerant. To this end, we design a family of quantum error-correcting codes on which sparse IQP circuits can be implemented in depth 1, meaning that they are fully parallelized.
This is possible thanks to the commuting nature of sparse IQP gates \cite{hoyer2005quantum}. In addition, we prove that the initial state can be encoded in constant quantum depth by performing stabilizer measurements. The only part of the process which is not implemented in constant depth is the final step of the state preparation: it consists of a single interaction with a classical computer that must compute a correction to apply, which depends on the stabilizer measurement results. In our scheme, this classical computation requires a polylogarithmic time because one needs to compute a correction for quantum patches of logarithmic size. We remark that similar time complexities are often neglected in the literature of quantum fault-tolerance~\cite{gottesman2014faulttolerant,aharonov1997fault,knill1998resilient}, and it may in fact not be a very problematic issue in practice.

\begin{figure}[t]
	\centering
	\includegraphics[width=0.46\textwidth]{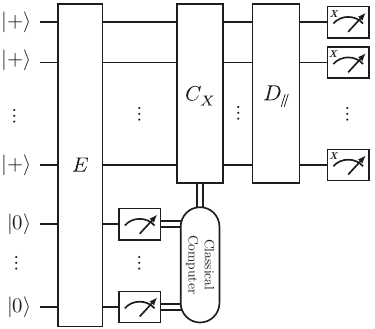}
	\caption{Our fault-tolerant implementation of a logical sparse IQP circuit $D$. The first layers $E$ (stabilizer measurements) and $C_X$ (adaptative error correction) prepare a logical state that is fed to a parallel version $D_{\varparallel}$ of the sparse IQP circuit $D$, followed by single-qubit measurements. The overall circuit has constant (quantum) depth and acts on $N \times \mathrm{polylog}(N)$ qubits. A single interaction with a classical computer is necessary to compute the correction $C_X$ for the initial preparation. A final classical post-processing (not depicted) then computes a sample from the target distribution $p_D$.}
	\label{fig:PIQP}
\end{figure}

Given a circuit $D \in \mathfrak{D}_N$ and precision $\delta > 0$, we construct the circuit $C_D(\delta)$ that samples from a distribution that is $\delta$-close to $p_D$ in total variation distance after classical post-processing. While the final classical post-processing is not performed in constant time, this is not an issue since all the qubits have already been measured. We discuss this point in Section~\ref{section:ft}. The circuit $C_D(\delta)$ is illustrated in Figure \ref{fig:PIQP} and we detail its construction in Section~\ref{section:thc}. 
For circuits $D$ of depth $\Theta(\log N)$, the space overhead is polylogarithmic in the precision $\delta$ and in number $N$ of logical qubits. Given that the average depth of sparse IQP circuits is $\Theta(\log N)$, a simple Markov inequality further implies that the fraction of circuits admiting a depth larger than $\alpha\log N$ decreases as $\order{1/\alpha}$. Thus an arbitrarily large fraction of such circuits benefits from the above overhead scaling. A practical difficulty that we do not address here is that our scheme requires long-range interactions. We state our main result:
\begin{theorem}[Constant depth quantum advantage] \label{th:qa}
	There exists a universal $\varepsilon_{\mathrm{th}} > 0$ such that, for all \(N\in\mathds
	N\), $D \in \mathfrak{D}_N$ and $\delta>0$, running a noisy version of the quantum circuit $C_{D}(\delta)$ by inserting local stochastic noise of strength \(\varepsilon < \varepsilon_{\mathrm{th}}\) after each step, yields samples from $p_D$ up to precision $\delta$ (in total variation distance) after classical post-processing.
\end{theorem}

Combining this with Theorem \ref{th:ch}, our scheme demonstrates a super-polynomial quantum advantage for the task of sparse IQP sampling, assuming Conjecture \ref{cj} and that the Polynomial Hierarchy does not collapse.

To summarize our contribution, we reduce the fault-tolerance space overhead required to demonstrate a superpolynomial quantum advantage with a constant depth quantum circuit, from a large degree polynomial in~\cite{mezher2020fault} to a polylogarithmic overhead. A similar reduction is achieved for the classical computation complexity during the quantum computation. This comes at the cost of losing the local connectivity of the scheme.

\subsection{Main concepts and ideas}

\subsubsection{The Tetrahelix code for fault-tolerant parallel computation \label{sssec:thc}}

We recall that we aim to address two issues in order to get a final circuit of constant depth: we need to reduce the depth of the logical circuit for sparse IQP from logarithmic to constant, and we need to find a fault-tolerant version that remains of constant depth. 
We achieve this by combining two ideas. 

\begin{figure}[t]
	\centering
	\includegraphics[width=0.4\textwidth]{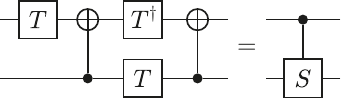}
	\caption{Implementation of the controlled-phase from controlled-not, $T$ and $T^\dagger$ gates, all of those have transversal implementation on a tetrahedral color code. Switching $T$ and $T^\dagger$ gives $CS^\dagger$ }\label{fig:CS}
\end{figure}

\begin{figure}[p]
	\centering
	\includegraphics[width=0.48\textwidth]{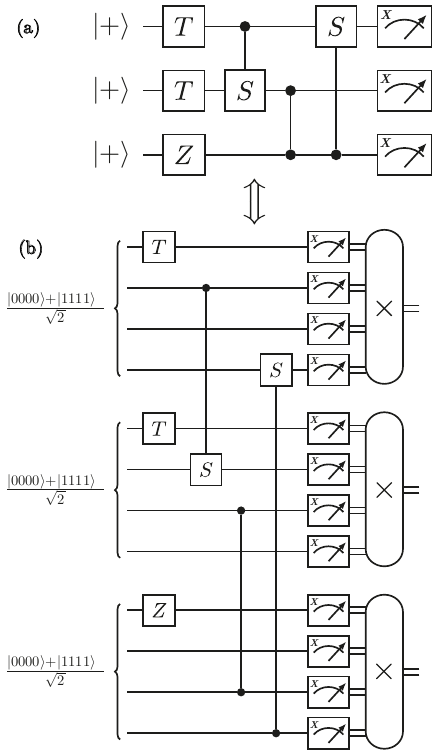}
	\caption{Each step $i \in \{1, ..., k\}$ of a circuit of depth $k$ is simultaneously applied on the $i^{th}$ physical qubit of all the GHZ states. The logical circuit (a) of depth 4 can be compiled in depth 1 up to classical decoding and state preparation by starting from GHZ state of size 4 and implementing circuit (b). The $\times$ blocks correspond to the classical decoding circuits.}
	\label{fig:Parallele}
\end{figure}

First we rely on \textit{3D color codes}~\cite{bombin2007topological,kubica2015universal} which admit transversal diagonal gates. More specifically, we will focus on the \textit{tetrahedral code} subfamily that admits a transversal $T$-gate. Moreover, because these codes are CSS codes~\cite{calderbank1996good,steane1996error}, they also admit a transversal CNOT gate. Combining both, we see that tetrahedral codes also have transversal $CS$-gates, as shown on Figure~\ref{fig:CS}.
The second idea is that it is possible to fully parallelize an IQP circuit by using a GHZ encoding of each of the input qubits, in order to trade depth for width of the circuit. This means encoding a $|+\rangle$ as $\frac{1}{\sqrt{2}} (|0\rangle^{\otimes k} + |1\rangle^{\otimes k})$ for some $k$ corresponding to the number of gates supposed to be applied to the qubit, so that $k$ is logarithmic in $N$.
Then all $k$ gates can be performed simultaneously by acting on a different qubit within the GHZ state. This is described in Figure \ref{fig:Parallele}. 
This new circuit has two shortcomings. First, despite being of constant depth, the logical phase-flip rate increases linearly with the size of the state since the measurement results of the $k$ qubits within a GHZ state need to be aggregated. Second the preparation of bare GHZ states cannot be done fault-tolerantly in constant depth.

To solve these issues, we define a new stabilizer code --  the \textit{tetrahelix code} --  combining the two ideas (3D color code for transversal gates and GHZ states for parallel implementation). We will detail the construction in Section~\ref{section:thc}, and only briefly explain its main properties here. 
The encoding is parameterized by two integers, $k$ and $L$, accounting respectively for the parallelization capacity and the distance of the code. A $k$-tetrahelix code of distance $L$ is defined by merging (in lattice surgery terms~\cite{horsman2012surface,litinski2019game,landahl2014quantum}) $k$ tetrahedral codes of distance $L$ along a 1-dimensional chain. Remarkably, the resulting $[\![\Theta(kL^3),1,\Theta(L)]\!]$ tetrahelix code admits a depth-1 implementation of a logical sparse IQP circuit of depth $k$. This corresponds to a linear trade-off between the depth of the initial logical circuit and the number of physical qubits:
\begin{lemma}\label{lemma:transversal}
	Any sparse IQP circuit of depth $k$ on $N$ qubits can be implemented in depth 1 on $N$ logical qubits encoded in $k$-tetrahelix codes.
\end{lemma}
The constant depth of the circuit, together with the arbitrarily large distance $L$, ensures the fault-tolerance of the circuit up to a final classical post-processing to decode the results of the qubit measurements. 
We discuss in Section~\ref{section:ft} how to achieve this by exploiting efficient decoders of color codes. The complexity of this step remains negligible compared to the super-polynomial quantum advantage of the overall circuit. 
The remaining challenge concerns the initial preparation of the encoded states of the tetrahelix code. 
One needs to ensure that such a preparation can also be done in constant depth and in a fault-tolerant manner.

\subsubsection{Single-shot state preparation}

A logical state of a quantum stabilizer code can always be prepared starting from a simple product state by measuring stabilizers and applying the appropriate correction to set the state in the code space. Such a scheme is however sensitive to measurement errors and fault-tolerance is usually achieved by repeating measurements. We circumvent this shortcoming by establishing the \textit{single-shot} preparation of logical $|+\rangle$ states for the tetrahelix code.

In general, error correction based on erroneous measurements can induce large-weight physical errors whose accumulation could later translate into logical errors. In order to ensure fault-tolerance, one can prove that building on the particular structure of syndromes, the induced residual errors can be kept local with high probability. Such errors are then dealt with by the final classical decoding step. This property corresponds to single-shot decoding introduced by Bombín in~\cite{bombin2015single}. Throughout this paper, $\ket{\overline x}$ denotes the logical encoded state $\ket{x}$ for $x \in \{0,1,+,-\}$.

\begin{lemma}\label{lemma:ss}
	The tetrahelix code admits a single-shot preparation of $\ket{\overline{+}}/\ket{\overline{-}}$ logical states, up to $X$ stabilizers of the tetrahedral code.
\end{lemma}

Note that, as argued in subsection \ref{ss:ssm}, the $X$ stabilizers need not be applied since they commute with sparse IQP encoded gates and hence can be propagated to the end of the circuit where they leave the final measurement unchanged.

The proof of the single-shot property of the $k$-tetrahelix code is detailed in Section~\ref{section:ss} and relies on ($i$) the single-shot preparation of Hadamard basis states for 3D gauge color codes~\cite{bombin2015single,bombin2015gauge}, and ($ii$) the fact that the measurement errors occurring during code merging are detectable with the global stabilizer measurement outcomes.

We furthermore argue that the associated decoding can be performed on a classical computer in polylogarithmic-time with respect to $N$ in Section~\ref{section:ss}. We consider it to be instantaneous to derive Theorem~\ref{th:qa}.

\subsection{Sketch of proof of Theorem~\ref{th:qa}}
The rest of the paper is devoted to establish Theorem~\ref{th:qa}.
In Section~\ref{section:thc}, we first briefly review tetrahedral codes, from the 3D color code family. Next, we define the tetrahelix code family obtained by merging tetrahedral codes. We prove that the $k$-tetrahelix code reduces the depth $k$ of a sparse IQP circuit to depth 1 (Lemma \ref{lemma:transversal}).
In Section~\ref{section:ss}, we prove that the merging failure probability between two tetrahedral codes of distance $L$ is exponentially suppressed in $L$ and hence that $k$-tetrahelix encoded states in the Hadamard basis can be faithfully prepared in constant quantum depth (Lemma \ref{lemma:ss}).
In Section~\ref{section:ft}, we prove the fault-tolerance of the scheme. More precisely, we prove the existence of a non-zero error threshold independent of $k$, below which we arbitrarily suppress logical errors by increasing $L$ for any encoded sparse IQP circuit.

\section{Tetrahelix code}\label{section:thc}
\subsection{Overview of tetrahedral codes}

\begin{figure}[t]
	\centering
	\includegraphics[width=0.46\textwidth]{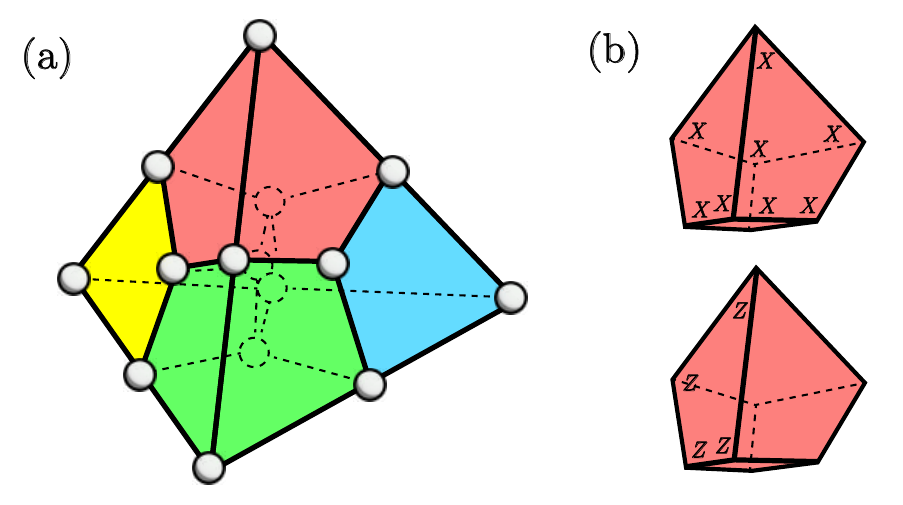}
	\caption{(a) The [15,1,3] Reed-Muller code is the smallest example of tetrahedral codes. Here qubits are on vertices and $\overline{X}$ and $\overline{Z}$ logical operators can be chosen respectively on a face and an edge of the tetrahedron. (b) $X$ stabilizers are supported on cells (elements of $\mathcal{C}$) and $Z$ stabilizers on faces (elements of $\mathcal{F}$).}
	\label{fig:ColorCode}
\end{figure}

\textit{Color codes} are a family of topological codes introduced by Bomb{\'\i}n and Martin-Delgado~\cite{bombin2007exact,bombin2007topological,kubica2015universal}. Their main feature is that they admit a transversal implementation of single-qubit phase gates, including the $T$-gate when the codes are 3-dimensional. In the following we focus on the subfamily of \textit{tetrahedral codes} that encode a single qubit.
Tetrahedral codes are defined on 3-dimensional color complexes, that we will call 3-colexes as in \cite{bombin2006topological}, of a tetrahedral shape as described in Figure \ref{fig:ColorCode} with the vertices corresponding to the data qubits.
3-Colexes are 3D lattices with the properties that ($i$) each cell is assigned one of four colors such that no two adjacent cells are of the same color; ($ii$) three colors appear on each external facet of the complex, and such a facet is associated with the missing color (in Figure \ref{fig:ColorCode} these facets correspond to the four triangular external boundaries of the tetrahedron); ($iii$) each vertex is incident to a cell or facet of all possible colors.

In the following we denote by $L \in \mathds{N}$ the number of vertices on the edges of the lattice, and will correspond to the code distance, as explained below. The construction of a tetrahedral 3-colex is not unique for a given $L$ but if one relies on tessellations of uniform density, then the resulting codes each encode a single logical qubit in $m=\Theta(L^3)$ physical qubits, and all display the properties that we will require. 

Let us recall the formal definition of a tetrahedral code on $m$ qubits, which will serve as a building block for the \textit{tetrahelix code}. We start from a tetrahedral 3-colex with set of vertices $\mathcal{V}$, faces $\mathcal{F}$ and cells $\mathcal{C}$. Physical qubits are associates with vertices, so $ m = |\mathcal{V}|$. The tetrahedral code associated to this colex is a CSS code with stabilizers given by:
\begin{equation}
	S_X^1 = \langle X(c), \: c \in \mathcal{C} \rangle,
\end{equation}
\begin{equation}
	S_Z^1 = \langle Z(f), \: f \in \mathcal{F} \rangle.
\end{equation}
Here, each cell $c$ or face $f$ is identified as a binary vector of length $m$ with ones at the locations corresponding to the associated vertices, and we define $X(a) := \otimes_{i=1}^m X_i^{a_i}$ for $a \in \{0,1\}^m$. (and similarly for $Z(a)$). In words, the $X$ stabilizer associated to a 3-cell $c$ is the product of Pauli $X$ operators on all the vertices in the boundary of $c$. In particular, $X$ stabilizers are associated by the 3-cells of the colex and $Z$ stabilizers are associated with the faces. 

The fundamental property of tetrahedral color codes is that for each code of the family there exists a partition of vertices $\mathcal{V}=\mathcal{V}^+ \cup \mathcal{V}^-$ such that applying the gate $T$ on $\mathcal{V}^+$ and $T^\dagger$ on $\mathcal{V}^-$ implements an encoded logical $T$-gate \cite{bombin2015gauge}:
\begin{equation}\label{eq:Ttrans}
	T(\mathcal{V}^+)T^\dagger(\mathcal{V}^-)\ket{\overline{x}}=\overline{T}\ket{\overline{x}}.
\end{equation}
Similarly to encoded states $\ket{x}$ denoted by $\ket{\overline{x}}$, we denote by $\overline{U}$ the encoded logical unitary $U$. Together with the existence of transversal controlled-not gates, this implies the transversal implementation of the $CS$-gate (see Figure \ref{fig:CS}):
\begin{equation}\label{eq:CStrans}
	CS(\mathcal{V}^+)CS^\dagger(\mathcal{V}^-)\ket{\overline{x}}\ket{\overline{y}}=\overline{CS}\ket{\overline{x}}\ket{\overline{y}},
\end{equation}
where $CS(\mathcal{V}^+)$ denotes the transversal application of $CS$ between the analogous sets $\mathcal{V}^+$ of two code blocks. $\overline{X}$ and $\overline{Z}$ logical operators are respectively surface-like and string-like and the $X$ distance and $Z$ distance scale as $\Theta(L^2)$ and $\Theta(L)$, respectively.

\subsection{Construction of the tetrahelix code} \label{Section_CTCC}

The transversality of the sparse IQP gate set paves the way towards the fault-tolerant implementation of such circuits. This would however require repeated error correction cycles at each circuit step, that is a logarithmic number of times. 
Concatenating a tetrahedral code with a repetition code gives a family of codes that present the desired parallelization property. Unfortunately, it does not meet the criteria of constant depth preparation for the initial encoded states. 
We now define a new code, the \emph{tetrahelix code}, that displays both properties: ($i$) depth-1 implementation of a sparse IQP circuit, ($ii$) constant-depth encoded state preparation in the Hadamard basis.

\begin{figure*}[t]
	\centering
	\includegraphics[width=0.93\textwidth]{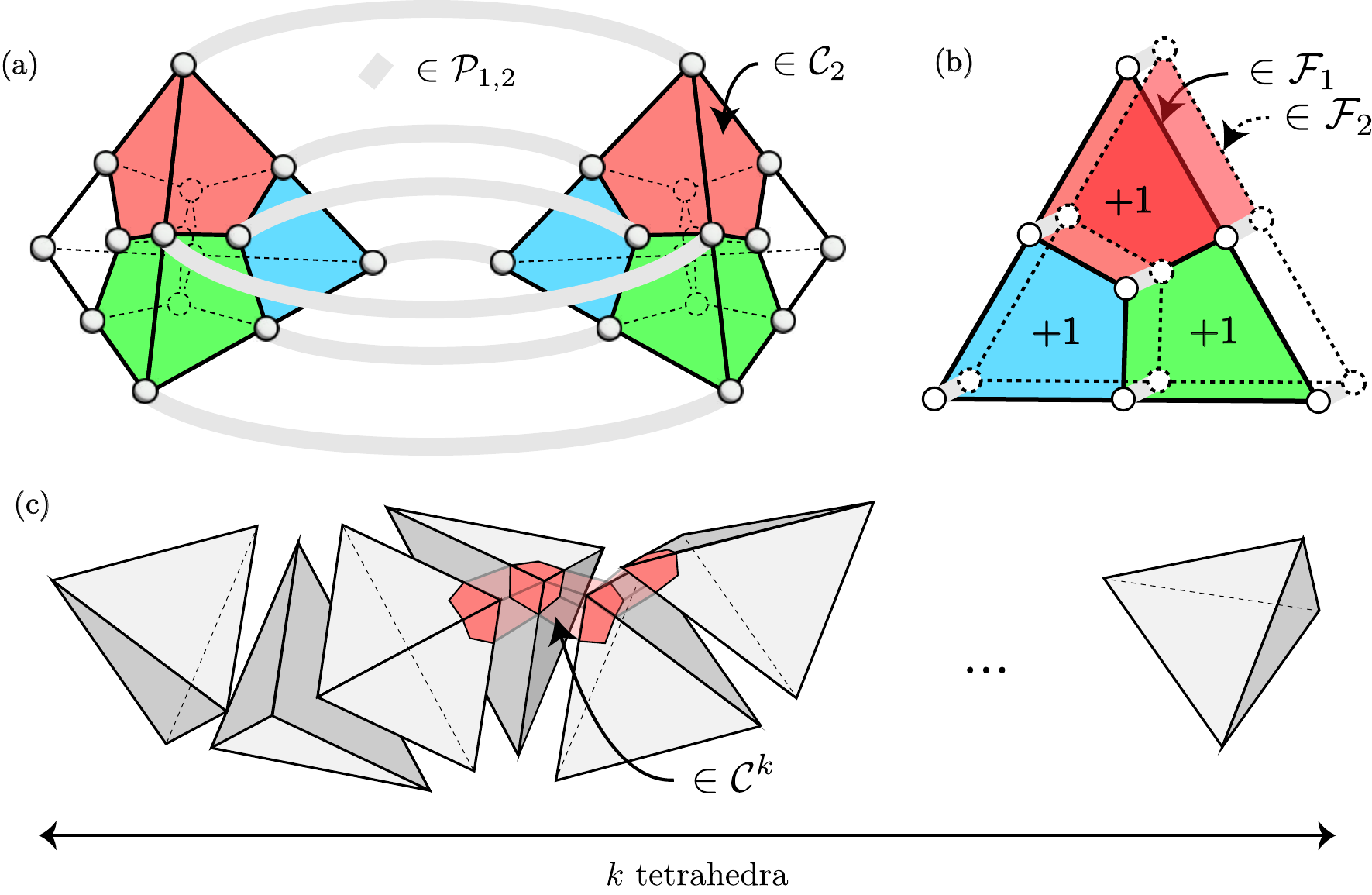}
	\caption{(a) Adjacent tetrahedra (here for $L=3$) are merged by measuring pairs of qubits from $\mathcal{P}_{1,2}$ that become $Z$ stabilizers of the new code. The colors of the second tetrahedron are chosen by convention so that merged $X$ stabilizers are of the same color. (b) Every new $Z$ stabilizer generator must be set to 1 to project the state in the code space. The new pair stabilizers value are not independent since they are all related to $Z$ stabilizers of the two tetrahedral codes on the face on which the merge is performed. In particular the product of four pairs overlapping the same $Z$ stabilizer (of support in $\mathcal{F}_1$) is equal to 1. (c) Merging additional tetrahedra with each other enables to form a chain of tetrahedra of length $k$. The optimal chain has the shape of an helix and corresponds to minimizing the number of merged $X$ stabilizers (of support in $\mathcal{C}^{k}$) that is equal to four in this packing.}
	\label{fig:Construction}
\end{figure*}

As briefly mentioned in subsection~\ref{sssec:thc}, a tetrahelix code is obtained by merging tetrahedral codes in lattice surgery terms~\cite{horsman2012surface,litinski2019game, vuillot2020fault}. Here we detail the construction starting by merging two such codes of distance $L$, with respective sets of vertices $\mathcal{V}_1$ and $\mathcal{V}_2$ as described in Figure \ref{fig:Construction}(a). We consider two codes which are exact mirror images of one another and we denote by \(\varphi:\mathcal{V}_1\rightarrow \mathcal{V}_2\) the bijection between the two sets of vertices.
An external triangular facet, \(B_1\subset \mathcal{V}_1\), together with its mirror image, \(B_2=\varphi(B_1)\), are chosen and every vertex is paired with the corresponding one on the other code. This pairing defines a set of pairs $\mathcal{P}_{1,2} := \left\{(v,\varphi(v))\vert \forall v\in B_1\right\}$.
\par The merge operation consists in fusing the $X$ stabilizers on the boundaries and adding new $Z$ stabilizers of weight 2 associated to the paired qubits in $\mathcal{P}_{1,2}$. More precisely, the $Z$ stabilizers are defined as $\langle Z(f),f\in\mathcal{F}^2\rangle$ with
\begin{equation}
	\mathcal{F}^2 := \mathcal{F}_1 \cup \mathcal{F}_{2} \cup \mathcal{P}_{1,2}.
\end{equation}
Similarly, we define $\mathcal{C}^2$ the union of merged stabilizers and unmerged ones:
\begin{equation}
	\mathcal{C}^2 := \mathcal{C}_{1}^* \cup \mathcal{C}_{2}^* \cup \mathcal{M}_{1,2}
\end{equation}
with
\begin{equation}
	\mathcal{C}_{i}^* := \{C \in \mathcal{C}_i | C \cap B_i = \varnothing \},
\end{equation}
and
\begin{equation}
	\begin{aligned}
		\mathcal{M}_{1,2} := \{ C_1 &\cup C_2 |  C_1 \in \mathcal{C}_1, C_2 \in \mathcal{C}_2,\\ &  \varphi(C_1 \cap B_1) = C_2\cap B_2 \ne \varnothing\}.
	\end{aligned}
\end{equation}
$X$ stabilizers are then defined as $\langle X(c),c\in\mathcal{C}^2 \rangle$ corresponding to non-adjacent cells of each code and fused adjacent ones (paired according to their color as in Figure~\ref{fig:Construction}(a)). This construction ensures that $X$ stabilizers commute with every $Z$ stabilizer including the newly defined $Z$ stabilizers with support on $\mathcal{P}_{1,2}$. We denote by $\overline{X}_i$ and $\overline{Z}_i$ the logical operators of the initial tetrahedral codes from which we define the logical operators of the new code. The 2-tetrahelix code encodes a single logical qubit for which the logical operators can be taken of the form $\overline{Z}=\overline{Z}_1$ (or $\overline Z_2$) and $\overline{X}=\overline{X}_1\overline{X}_2$ with $\overline{X}_2$ chosen so that its support intersects on the same subset of $\mathcal{P}_{1,2}$ than $\overline{X}_1$ (to commute with the associated $Z$ stabilizers).

Merging additional tetrahedra does not fundamentally change the analysis. Tetrahedra can be aligned in the shape of Figure \ref{fig:Construction}(c) to form a chain of length $k \in \mathds{N}$ so that each extremal vertex is shared between at most four tetrahedra. This ensures that at most four $X$ stabilizers are fused together. This linear packing of regular tetrahedra is known as a Boerdijk-Coxeter helix or tetrahelix~\cite{boerdijk1952some,coxeter1994regular} which motivates the name of the code.

Denoting by $\mathcal{V}_i$ the set of vertices of the $i^{th}$ tetrahedral color code, we get a partition of the set of all vertices $\mathcal{V}=\cup_{i=1}^k \mathcal{V}_i$. With $\mathcal{P}_{i,i+1}$ denoting new $Z$ stabilizers between adjacent tetrahedra $i$ and $i+1$, $\mathcal{F}^k$ and $\mathcal{C}^k$ are defined analogously as $\mathcal{F}^2$, and $\mathcal{C}^2$ to ensure stabilizer commutation:
\begin{equation}
	\mathcal{F}^k := \bigcup_{i=1}^{k-1}(\mathcal{F}_i \cup \mathcal{P}_{i,i+1}) \cup \mathcal{F}_k,
\end{equation}
\begin{equation}
	\mathcal{C}^k := \bigcup_{i=1}^{k-1} (\mathcal{C}_i^* \cup \mathcal{M}_{i,i+1})\cup \mathcal{C}_k^*,
\end{equation}
where here $\mathcal{C}_{i}^*$ and $\mathcal{M}_{i,i+1}$ are defined recursively so that stabilizers can be merged across several tetrahedra (up to three on edges and up to four on summits). We define the $k$-tetrahelix code that encodes a single logical qubit in $\Theta(kL^3)$ physical qubits from its set of stabilizers:
\begin{equation}
	S_X^k := \langle X(c), \: c \in \mathcal{C}^k \rangle,
\end{equation}
\begin{equation}
	S_Z^k := \langle X(f), \: f \in \mathcal{F}^k \rangle.
\end{equation}

The logical $\overline{Z}$ operator can be chosen as any of the logical $\overline{Z}_i$ operators of the composing tetrahedral codes, while the $\overline X$ logical operator is a product of $\overline{X}_i$ operators recursively chosen so that $\overline{X}_i$ and $\overline{X}_{i+1}$ intersect with the same subset of $\mathcal{P}_{i,i+1}$.

\subsection{Code distance}
\par The $X$ and $Z$ distances of a code correspond to the minimal weights of $X$ and $Z$ logical operators. We denote by $d_X^k$ and $d_Z^k$ the $X$ and $Z$ distances of the $k$-tetrahelix code and prove that:
\begin{equation}\label{eqn:dist}
	d_Z^k = \Theta(L) \quad \text{and} \quad d_X^k = \Theta (kL^2).
\end{equation}
We prove the result for the 2-tetrahelix code by relating logical operators of the tetrahelix code to those of the initial tetrahedral codes, the result generalizes to arbitrary $k$-tetrahelix code by recursion. In this subsection we denote by $d_X^1 = \Theta (L^2)$ and $d_Z^1=\Theta(L)$ the $X$ and $Z$ distances of a tetrahedral code of edge length $L$ and number of qubits $m=\Theta(L^3)$.

\par Let us consider a logical operator $\overline{Z}$ of the 2-tetrahelix code. We index the vertices of the two composing tetrahedra in a symmetric manner with respect to the paired facets. The logical operator $\overline Z$ is of the form of the tensor product of Pauli $Z$ operators on each tetrahedron $\overline{Z}=Z(\mu) \otimes Z(\nu)$, 
with $\mu,\nu\in\{0,1\}^m$. We will show that up to multiplication by $Z$ stabilizers we can transfer $Z(\mu) \otimes Z(\nu)$ to $Z(\mu+\nu)\otimes \mathds{1}$. This means that we transfer the physical Pauli $Z$ operators from the second tetrahedron to the symmetric ones in the first one. We can then conclude by noticing that $Z(\mu+\nu)$ is a logical operator of the first tetrahedron and hence of weight larger or equal to $d_Z^1$. We thus have
\begin{equation}
	|\overline{Z}|=|Z(\mu) \otimes Z(\nu)|  \geq |\overline{Z}(\mu + \nu) \otimes \mathds{1}| \geq 
	d_Z^1 ,
\end{equation}
which concludes the argument.
\par Indeed, since an arbitrary logical $\overline{Z}_1$ operator of the first tetrahedron is also a logical operator of the tetrahelix code, there exists a $Z$ stabilizer $R_Z$ such that $\overline Z=\overline Z_1\times R_Z$. Such a stabilizer is necessarily of the form:
\begin{equation}
	R_Z=R_Z^1\times R_Z^2\times R_Z^{1,2},
\end{equation}
where $R_Z^1$ is a $Z$ stabilizer of tetrahedron 1, $R_Z^2$ of tetrahedron 2, and $R_Z^{1,2}$ is a product of Z-stabilizers defined at the boundary (paired qubits in $\mathcal{P}_{1,2}$). It is clear that, multiplying $\overline Z$ by $R_Z^2$, maps the support from the bulk of the second tetrahedron to the paired facet of this tetrahedron. Next, we completely transfer this support to the facet of first tetrahedron by multiplying by $R_Z^{1,2}$. At this point, the support of the logical operator is entirely contained in the first tetrahedron. Now we apply the symmetric version of $R_Z^2$ defined on the first tetrahedron. This maps the original logical operator to $Z(\mu+\nu)\otimes \mathds{1}$.

The case of the $X$ distance is straightforward as the product of $\overline{X}_1$ and $\overline{X}_2$ logical operators whose supports intersect with the same pairs of $\mathcal{P}_{1,2}$ yields a logical operator of the 2-tetrahelix code, and this product form is stable upon multiplication by $X$ stabilizers. This stability is a direct consequence of the fact that the restriction of a tetrahelix $X$ stabilizer to a single tetrahedron is a stabilizer of the tetrahedral code.
Merging an additional tetrahedral code hence increases the $X$ distance by $d_X^1$:
\begin{equation}
	|\overline{X}|=|\overline{X}_1\overline{X}_2| \geq 2 d_X^1.
\end{equation}
The same discussion between a $(k-1)$-tetrahelix code and a tetrahedral code generalises the proof by recursion to $k$-tetrahelix code for arbitrary $k$. Recalling that $d_Z^1 = \Theta(L)$ and $d_X^1 = \Theta(L^2)$, we obtain the bounds of \eqref{eqn:dist}.

\subsection{Parallel computation}
\label{ss:pc}
We turn to the properties of the code concerning parallel computation.
We establish Lemma \ref{lemma:transversal} by showing that the encoded $T$-gate can be implemented in depth 1 on a single tetrahedron of the chain. A tetrahedral code on $m$ physical qubits is a CSS code and logical states can therefore be written in the form
\begin{equation}\label{eq:ecritureCSS}
	\ket{\overline{x}_1} = \frac{1}{\sqrt{|\mathcal{S}^1|}} \sum_{s_1 \in \mathcal{S}^1} \ket{s_1 + x_1L_1},
\end{equation}
with the addition taken modulo 2 and $x_1 \in \{0,1\}$ and $\mathcal{S}^1 \subset \{0,1\}^m$ such that for $s_1 \in \mathcal{S}^1$ we have $X(s_1) \in S_{X}^1$.
Similarly, $L_1 \in \{0,1\}^{m}$ represents an arbitrary $\overline{X}_1$ logical operator. The transversal implementation of the $T$-gate $T(\mathcal{V}^+)T^\dagger(\mathcal{V}^-)=\overline{T}$ on the tetrahedral code implies that each codeword gains the same phase from the application of $T(\mathcal{V}^+)T^\dagger(\mathcal{V}^-)$:
\begin{equation}\label{eq:TtransTH}
	T(\mathcal{V}^+)T^\dagger(\mathcal{V}^-)\ket{s_{1} + x_1L_{1}} = e^{\frac{i\pi}{4}\abs{x_1}}\ket{s_{1} + x_1L_{1}}.
\end{equation}
The logical computational states of the $k$-tetrahelix code are given by
\begin{equation}
	\ket{\overline{x}} = \frac{1}{\sqrt{|\mathcal{S}^k}|} \sum_{s \in \mathcal{S}^k} \ket{s + xL},
\end{equation}
for $x \in \{0,1\}$. Here $\mathcal{S}^k\subset\{0,1\}^{k\times m}$ is such that for $s\in\mathcal{S}^k$, we have $X(s)\in S_X^k$ and $L$ is the vector associated to an arbitrary logical $\overline X$ operator. For each $X$ stabilizer, $s\in\mathcal{S}^k$ is a concatenation of $k$ vectors $s_i \in \mathcal{S}^1, i \in \{1,\dots,k\}$: $s = [s_1, ..., s_k]$. Similarly, for the logical operator, the binary vector $L$ is a concatenation of vectors $L_i$ each representing a logical $\overline X_i$ operator of the $i^{th}$ tetrahedral code. Focusing on tetrahedron $i_0$, and up to qubit re-ordering, we can thus write
\begin{equation}\label{eq:Decompo}
	\ket{\overline{x}} = \frac{1}{\sqrt{|\mathcal{S}^k}|} \sum_{s_{i_0} \in \mathcal{S}^1} \ket{s_{i_0} + xL_{i_0}} \otimes \ket{\psi(s_{i_0},x)}.
\end{equation}
The terms $\ket{\psi(s_{i_0},x)}$ depend on $s_{i_0}$ because of correlations between codewords restricted to different tetrahedra induced by overlapping $X$ stabilizers, but this does not impact our argument. Taking $\mathcal{V}^+_{i_0}$ and $\mathcal{V}^-_{i_0}$ as in~\eqref{eq:Ttrans}, we have
\begin{equation}\label{eq:CW}
	T(\mathcal{V}^+_{i_0})T^\dagger(\mathcal{V}^-_{i_0})\ket{s_{i_0} + xL_{i_0}} = e^{\frac{i\pi}{4}\abs{x}}\ket{s_{i_0} + xL_{i_0}}.
\end{equation}
Combining~\eqref{eq:Decompo} and~\eqref{eq:CW} directly implies:
\begin{equation}
	T(\mathcal{V}^+_{i_0})T^\dagger(\mathcal{V}^-_{i_0})\ket{\overline{x}} = \overline{T}\ket{\overline{x}}.
\end{equation}
To extend the arguments to the $CS$-gate, we would need to use the gadget of Figure~\ref{fig:CS}. Notice that while the $T$ and $T^\dag$-gates are applied on a single tetrahedron, the CNOT gates would need to be applied on all physical qubits (CSS property). However, as these CNOT gates come in pairs, they cancel each other outside the tetrahedron where the $T$-gates are applied. Thus, the $CS$-gate can also be applied between two arbitrary tetrahedral blocks of two tetrahelix codes:
\begin{equation}
	CS(\mathcal{V}_{i_0}^+)CS^\dagger(\mathcal{V}_{i_0}^-)\ket{\overline{x}}\ket{\overline{y}}=\overline{CS}\ket{\overline{x}}\ket{\overline{y}}.
\end{equation}
This implies that the $k$-tetrahelix code can implement in a single step an encoded $T$ or a $CS$-gate on each tetrahedral block of the code. Therefore, we obtain a depth-1 parallel implementation of a depth-$k$ sparse IQP circuit up to state preparation. This finishes the proof of Lemma~\ref{lemma:transversal}. In the next section we prove that encoded states in the Hadamard basis can be prepared fault-tolerantly in constant quantum depth.

\section{Constant-depth preparation of encoded states}\label{section:ss}

The encoded states of the $k$-tetrahelix code in the Hadamard basis can be prepared by merging $k$ associated encoded states of the composing tetrahedral blocks. In this section, we show that both the steps of preparing encoded tetrahedral states and their merging can be done in constant quantum depth.

\subsection{Single-shot decoding of tetrahedral code}

Since the state $\ket{+}^{\otimes m}$ is stabilized by all $X$ stabilizers and by the $\overline{X}$ logical operator of a CSS quantum code, the projection over the logical $\ket{\overline{+}}$ can ideally be done by measuring the $Z$ stabilizers and a single step of Pauli $X$ corrections. Measurement errors however usually prevent such reliable encoded state preparation in constant depth. Indeed, measurement errors induce residual data errors after Pauli corrections, which usually calls for many repeated measurements before such a correction is applied. Repeating measurements gives an extra dimension to the error syndrome where ancilla and data errors can be separated by the decoding algorithm which infers an error pattern close to the most likely one and hence an appropriate correction.

An alternative approach is to build on the structure of the error syndrome of some codes to ensure that a single round of local measurements is sufficient to ensure the locality of the residual errors with high probability. This strategy was first proposed by Bomb{\'\i}n in \cite{bombin2015single} for 3D gauge color codes and is known as \textit{single-shot decoding} which is a property of a quantum error-correcting code in conjunction with its decoder. In the case of 3D color codes, $Z$ stabilizers on faces correspond to $Z$ gauge operators of 3D gauge color codes. This ensures the single-shot decoding property up to a classical computation of polynomial complexity in the code size. Furthermore, the topological nature of the code implies that the measurements can be parallelized to a constant quantum depth. In conclusion, we have a constant depth preparation of encoded states in the Hadamard basis for the tetrahedral code up to local residual errors.

\subsection{Single-shot merging of tetrahedral states} \label{ss:ssm}

Merging two tetrahedral codes of distance $L$ into a 2-tetrahelix code is described in Figure \ref{fig:Construction}(a-b). Pairs of qubits from $\mathcal{P}_{1,2}$ are measured over the faces on which tetrahedral codes are merged and a correction is applied depending on the measurement outcomes. Since facets of a tetrahedral code have the structure of a triangular code (2D color code) of size $L$, in the absence of errors, measurements yield a binary codeword $w$ of the corresponding classical code. This codeword can be written as the sum of an $X$ stabilizer and an $X$ logical operator of the 2D code with the same formalism used in Section \ref{ss:pc} for the 3D code.

The appropriate correction then can be seen to be a 3D color code codeword whose restriction to the triangular code vertices gives $w$. This can be obtained by determining first the decomposition of $w$ into facets of the triangular code ($X$ stabilizer generators) and the logical operator $X$ over the entire triangle (so that it is a logical operator of both the 2D and the 3D codes) before mapping the facets to cells to get a 3D code codeword

\begin{align}\label{eq:corr}
\begin{split}
    w = s_{2D} & + xL_{2D} \\
    & \rightarrow s_{3D} + xL_{3D} = \text{corr}(w)
\end{split}
\end{align}
for $x \in \{0,1\}$. Importantly, an $X$ stabilizer of the tetrahedral code commutes with encoded $T$ and $CS$-gates on the tetrahelix code since it does not change the structure of the codewords described in subsection \ref{ss:pc}. Since the circuit ends with $X$ measurements this means that it is sufficient for our purpose to compute $x$ and only apply the logical part of the correction. In other words, we only need to prepare tetrahelix encoded states up to tetrahedral codes $X$ stabilizers.

If the two tetrahedral codes are not perfectly in their code space, or in the case of measurement errors, the measurement results deviate from $w$:

\begin{equation}
	w_e = w + e_r + e_m.
\end{equation}

Here, $e_r$ accounts for the residual errors of the tetrahedral states preparation, and $e_m$ stands for measurement errors. Because the preparation of encoded states in the Hadamard basis for the tetrahedral code is single-shot, the resulting errors follow a local stochastic noise model. This is also the case for measurement errors and hence decoding the triangular code yields the correct value of $\overline{Z}_1\overline{Z}_2$ with probability exponentially close to 1 in $L$. $\overline{Z}_1\overline{Z}_2$ is then set to +1 by applying or not $\overline{X}_1$.

A $k$-tetrahelix code encoded state can then be prepared in a similar manner simply by repeating the merging operation with additional tetrahedral codes, while always applying the logical correction on the tetrahedron for example on the left of the merge. This scheme can be seen as similar to preparing a GHZ state of size $k$ from parity measurements and logical correction, with the difference that here, measurement errors are exponentially suppressed, thus giving Lemma \ref{lemma:ss}.

Efficient decoding algorithms exist for 2D color codes~\cite{chamberland2020triangular, sahay2022decoder} and 3D color codes~ \cite{bombin2015single,kubica2023efficient} and are single-shot for the 3D case. These algorithms have a complexity polynomial in the code size. 
The different tetrahedral encoded states can be prepared in parallel. Parallel merge measurements followed by iterative computation of the associated correction then give a preparation of tetrahelix encoded states with polynomial in $L$ and proportional to $k$ classical computation. We prove in Section~\ref{section:ft} that we need a polylogarithmic number of qubits per code block which hence gives a polylogarithmic-time classical computation.

\section{Application to sparse IQP circuits}\label{section:ft}

In this section, we apply the results of the two previous sections to demonstrate the main result of this paper stated in Theorem~\ref{th:qa}. We start by presenting the error model. Next, we show that the encoding of the circuit of Figure~\ref{fig:PIQP} is fault-tolerant by proving the existence of an error threshold. Finally, we provide an estimation of the space overhead of the scheme.

\subsection{Error model}

The coupling of the quantum system with the environment generates noise that can later induce errors in the computation. We use the \textit{local stochastic quantum noise model} from \cite{gottesman2014faulttolerant} where the set of faulty locations is a random variable of a discrete space-time and local correlations are allowed. No assumption is made on a particular type of error operator. This makes the model general enough to cover a wide class of applications. In particular this captures commonly studied noise channels such as depolarizing and dephasing noise, or amplitude damping.

A noise model of parameter $\varepsilon$ that satisfies the following two properties is said to be locally stochastic: ($i$) the faults are confined to a random set of space-time locations $\mathcal{A} \subset V$ with probability $p(\mathcal{A})$ and ($ii$) the probability that a set of faulty locations contains a specific set of $\mathcal{A}$ locations is upper bounded by $\varepsilon^{|\mathcal{A}|}$.
\begin{equation}
	\sum_{\mathcal{A}' \supseteq \mathcal{A}} p(\mathcal{A}') \leq \varepsilon^{|\mathcal{A}|}.
\end{equation}
Final measurements are performed in the Hadamard basis and hence at the end of the circuit only $Z$-type errors induce errors on the classical output. $Z$ errors can either be environmentally induced or generated during the propagation of $X$-type errors in the circuit. $X$ errors can also arise due to the coupling with the environment but also from incorrect preparation of encoded states (recall that the preparation only includes $X$ correction). Local stochastic errors propagate as such through the constant depth circuit but residual errors after encoded states preparation are not necessarily local. We showed in Section~\ref{section:ss} that their non-local representatives admit exponentially low probabilities which implies that the correction of local stochastic errors by the final decoding is sufficient to exponentially suppress the logical error rate.

More formally, residual errors after preparation of tetrahedral encoded states are characterised in~\cite{bombin2015single} such that ($i$) correctable physical errors follow a local stochastic noise model $\mathcal{N}_{\text{T,loc}}^\varepsilon$, ($ii$) non-correctable physical errors (that is to say errors whose correction attempt induces a logical error) are exponentially suppressed, we denote by $\mathcal{N}_{\text{T,nc}}^{\tilde{\varepsilon}_1}$ the corresponding error channel. Using a similar notation we call $\mathcal{N}_{\text{M,nc}}^{\tilde{\varepsilon}_2}$ the channel associated to logical errors due to unsuccessful merging, with probability exponentially suppressed in the code distance.

The encoded states preparation error channel then writes with $\tilde{\varepsilon}_1$ and $\tilde{\varepsilon}_2$ exponentially suppressed in $\varepsilon$:

\begin{equation}
	\mathcal{N}^\varepsilon_{\text{prep}} = \mathcal{N}_{\text{T,loc}}^\varepsilon \circ \mathcal{N}_{\text{T,nc}}^{\tilde{\varepsilon}_1} \circ \mathcal{N}_{\text{M,nc}}^{\tilde{\varepsilon}_2}.
\end{equation}
The two non-correctable terms contribute to the final logical error rate but are exponentially rare. In the following subsections we prove that low enough local stochastic noise is corrected by the tetrahelix code. Post-processing of final single qubit measurements in the form of the tetrahelix code decoding then yields the value of the logical measurement. In the following we analyze error configurations and describe an efficient decoder from 2D and 3D color code decoders.

\subsection{Existence of a good decoder}

\begin{figure}[t]
	\centering
	\includegraphics[width=0.46\textwidth]{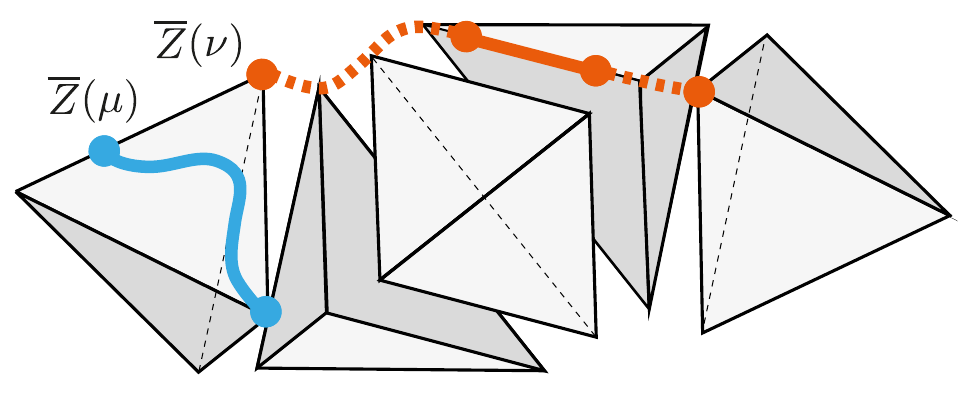}
	\caption{Representation of $\overline{Z}$ logical error configurations. Error strings are no longer restricted to a single tetrahedron (blue) but can also connect neighbouring tetrahedra (red). In the tetrahelix stacking, error strings can jump up to two tetrahedra at once.}
	\label{fig:errors}
\end{figure}

In the 3D color code, the logical $\overline{Z}$ operator corresponds to strings of Pauli Z connecting the four boundaries of different colors. An extremal vertex of the tetrahedron belongs to three boundaries and a string connecting this vertex to the opposite face of the remaining color hence yields an example of a $\overline{Z}$ logical operator. Logical errors arise when more than half of the respective phases of any such path are flipped. Errors on the tetrahelix code have a similar origin except that error strings can jump between tetrahedra to connect boundaries of different colors as described in Figure \ref{fig:errors}. This means that we cannot individually decode tetrahedra and that we first need to split (in lattice surgery language) the chain to retrieve tetrahedral codes.

This can be performed in software after final single-qubit $X$ measurements by reconstructing the value of $X$ stabilizers of the tetrahedral codes. Considering the example of the 2-tetrahelix code for simplicity, $X$ stabilizers at the interface between the two tetrahedra were merged and hence, taken individually, do not stabilize the tetrahelix code. This means that they will initially not necessarily be in their +1 eigenspace even without errors. This can be fixed by applying $Z$ stabilizers from $\mathcal{P}_{1,2}$ to set them to +1 while acting trivially on the code space of the tetrahelix code. This can be seen as preparing two triangular codes (2D color codes) logical states on the two facets by applying a Pauli operator of the form

\begin{equation}
    Z(\sigma)\otimes Z(\varphi(\sigma)),
\end{equation}
with $\sigma \subset B_1$ a set of vertices from the triangular facet on which the merge was performed, and $\varphi$ the bijection between the two tetrahedra sets of vertices defined in Section~\ref{section:thc}. Note here that any potential logical error applied to one tetrahedron would also be applied to the second one.

In reality, errors arising on the support of tetrahedral codes $X$ stabilizers prevents all such stabilizer to be set back to +1 by applying $Z$ stabilizers from $\mathcal{P}_{1,2}$. Since in this scheme we aim at correcting errors at the next step during individual tetrahedral codes decoding we only need here to approach the tetrahedral code spaces. This can be done by minimizing the number of tetrahedral code $X$ stabilizers with value -1 in the chain. For the $k$-tetrahelix code we start by $X$ stabilizers merged between more than two tetrahedra, that is to say on tetrahedra vertices and edges, followed by those on the bulk of the facet on which tetrahedral codes are merged.

Once each tetrahedron is back to the tetrahedral code space (up to physical errors) it suffices to individually decode each code and multiply the logical values of the $\overline{X}_i$'s to recover the desired logical information (and hence pairs of tetrahedral codes logical errors possibly introduced at the splitting step cancel each other). For a low enough error rate we thus expect the logical error rate $\overline{\varepsilon}$ after such decoding to be proportional to the number of tetrahedra in the chain and to the logical error rate of a single tetrahedral code:
\begin{equation}
	\overline{\varepsilon} = \order{kL^3}(\varepsilon/\varepsilon_{\mathrm{th}})^{\order{L}}.
\end{equation}
Here we have only used 2D and 3D color codes decoders and therefore the existence of efficient 2D and 3D color code decoders \cite{chamberland2020triangular,beverland2021cost,sahay2022decoder,kubica2023efficient} implies the existence of an efficient decoder for the tetrahelix code. The formal definition and analysis of such a decoder under the general noise model considered here is beyond the scope of this paper and in the following we will prove Theorem \ref{th:qa} by relying on existing results on quantum LDPC codes. To do so we show that the code admits a non-zero threshold independent of $k$ so that for low enough noise the logical error can be made arbitrarily low by increasing $L$.

\subsection{Existence of a threshold for minimum weight decoder and local stochastic noise}
The tetrahelix code is a quantum LDPC code since its generators have a bounded weight and each qubit is involved in a bounded number of generators. It is known that a family of \([[\tilde{n},\tilde{k},\tilde{d}]]\) quantum LDPC codes, with $\tilde n$ and $\tilde d$ scaling to infinity, and experiencing local stochastic noise of parameter $\varepsilon$, admits a non-zero error threshold $\varepsilon_{\mathrm{th}}$ \cite{gottesman2014faulttolerant}. More precisely, below this threshold the logical error rate is exponentially suppressed as
\begin{equation}\label{eq:gott}
	\overline{\varepsilon}=\order{\tilde{n}(\varepsilon/\varepsilon_{\mathrm{th}})^{\tilde{d}/2}},
\end{equation}
using the \textit{minimum weight decoder}.

In the case of tetrahelix code, $k$ is in general an independent parameter from $L$ the distance of the code. Applying directly the results of~\cite{gottesman2014faulttolerant} would lead to a threshold dependent on the value of $k$. We take care of this issue by imposing $k=\order{L}$. Thus, the associated family of $k$-tetrahelix codes admits a number of physical qubits $\tilde n=\order{L^4}$, and such a \( [[\tilde{n}=\Theta(kL^3),\tilde{k}=1,\tilde{d}=\Theta(L)]]\) code admits a non-zero threshold $\varepsilon_{\mathrm{th}}$ with $L$ scaling to infinity. Note that, while the minimum weight decoder is not efficient in general, we expect the efficient decoder of the previous subsection to present similar error suppression property.

In the following subsection, we show that imposing $k=\order{L}$ is compatible with the desired parallelization and fault-tolerance properties.

\subsection{Proof of Theorem \ref{th:qa}}

When implementing sparse IQP circuits on $N$ qubits on $k$-tetrahelix codes, $k$, $L$ and $N$ are related through three relations. First, as discussed in the previous subsection, to ensure the existence of threshold independent of $k$, we need to have 
\begin{equation}\label{eq:relation}
	k = \order{L}.
\end{equation}
Second, an arbitrarily large fraction of sparse IQP circuits on $N$ qubits are of depth $\Theta(\log N)$ and can hence be implemented on $k$-tetrahelix codes with:
\begin{equation}\label{eq:c1}
	k=\Theta(\log N),
\end{equation}
Third, another relation between $L$ and $N$ results from the required code size to reach the target precision $\delta$ of the sparse IQP problem. A logical sparse IQP circuit $D$ is implemented with a $k$-tetrahelix code by the circuit $C_D$. After the final decoding, one obtains samples from a distribution $p_{D,\overline{\varepsilon}}$. For a constant logical error rate $\overline{\varepsilon}$ per logical qubit, the union bound gives an upper bound to the distance between the noisy and the ideal probability distributions with respect to $N$:
\begin{equation}\label{eq:ub}
	\norm{p_D-p_{D,\overline{\varepsilon}}}_{\text{TV}}\leq \order{N \overline{\varepsilon}}.
\end{equation}
Keeping the noisy probability distribution $\delta$-close to the ideal distribution thus imposes a logical error rate $\overline{\varepsilon}$ at most $\order{\delta/N}$. Logical errors can arise both from local stochastic errors and remaining non-local residual errors induced by merging errors or tetrahedral encoded states preparation errors, all of which are exponentially suppressed in $L$ below some threshold:
\begin{equation}\label{eq:thresh}
	\overline{\varepsilon} = \left(\frac{\varepsilon}{\varepsilon_{\mathrm{th}}}\right)^{\Theta(L)},
\end{equation}
where the polynomial dependency on $L$ in equation~\eqref{eq:gott} is absorbed by the exponential. From \eqref{eq:ub} and \eqref{eq:thresh}, we derive the third equation relating $L$ and $N$,
\begin{equation}\label{eq:c2}
	L = \Omega\left(\frac{\log(N/\delta)}{\log(\varepsilon_{\mathrm{th}}/\varepsilon)}\right).
\end{equation}
For a given $N$, it is enough to take 
\begin{equation}
	L = \Theta\left(\frac{\log(N/\delta)}{\log(\varepsilon_{\mathrm{th}}/\varepsilon)}\right)\quad \text{and} \quad k=\Theta(\log N),
\end{equation}
which automatically also satisfy~\eqref{eq:c1}. The total number of qubits $n$ of each code block for a sparse IQP circuit of width $N$ then reduces to:
\begin{equation}
	n = \Theta(kL^3) = \Theta\left(\frac{\mathrm{polylog}(N/\delta)}{\mathrm{polylog}(\varepsilon_{\mathrm{th}}/\varepsilon)}\right).
\end{equation}
This completes the proof of Theorem \ref{th:qa}. We note that for the (arbitrarily small) fraction of sparse IQP circuits of super-logarithmic depth the overhead is at most polynomial since the depth of sparse IQP circuits is at most linear.

\section{Discussion}

We have proposed a fault-tolerant implementation of sparse IQP circuits, paving the way for demonstrations of super-polynomial quantum advantage in near or mid-term experiments.
It consists in a constant-depth quantum circuit and involves a single step of feed-forward from classical computation.
To do this we have introduced the tetrahelix code admitting single-shot preparation of logical \(\ket{+}\) states and transversal implementation of IQP circuits.
The qubit overhead and classical computation time of our scheme are only polylogarithmic in the width of the original sparse IQP circuit. 
The requirements of our protocol are almost met by current NISQ experiments.
We hope it can bring within reach demonstration of super-polynomial advantage of quantum over classical computation.

Depending on the physical platform the main complexity of the protocol may be coming from the required connectivity.
A single tetrahelix code requires 3D connectivity.
On top of this, each physical qubit has to interact with a single other qubit from another tetrahelix code during the implementation of the IQP circuit.
These additional interactions are potentially long range. In the same spirit as in~ \cite{litinski2019game},  the interaction range can be reduced using longer and branching tetrahelix codes while staying 3D.
Logical \(CZ\)-gates can be realized facet to facet \cite{bombin2018transversal}, but \(CS\)-gates will still require transversal connectivity between tetrahedra.
Finding other codes with similar properties but simpler connectivity could ease the implementation even more.

The tetrahelix code that we propose in our implementation has interesting properties in itself.
The ability to implement many different non-Clifford unitaries in a transversal manner could potentially be leveraged in other settings.
One can take inspiration from this construction to design other codes with large sets of transversal non-Clifford unitaries to locally trade depth for width in larger scale algorithms. The key ingredient of our approach is the commuting nature of sparse IQP gates that enables their parallelization, in the spirit of \cite{browne2011computational} for MBQC, but in a fault-tolerant manner. Note that a generalization of the construction to any set of commuting gate would have powerful applications \cite{hoyer2005quantum}.

Concerning the encoding of sparse IQP circuits on tetrahelix codes, it is not clear whether or not the trade-off between depth and width is optimal but the noticeable asymmetry between the \(X\) and \(Z\) distances of the tetrahelix code seems to indicate the overhead could be reduced by balancing them out. Another direction would be to improve the error threshold of the scheme, possibly with post-selection in the spirit of \cite{fujii2016noise}. This would participate bringing the scheme further within reach of current experiments \cite{bluvstein2024logical}.

During the preparation of this work, we became aware of a similar work on reducing error correction requirements for fault-tolerant quantum advantage \cite{Markham}.

\section*{Acknowledgements}

We acknowledge support from the Plan France 2030 through the project NISQ2LSQ ANR-22-PETQ-0006, HQI ANR-22-PNCQ-0002 and from Inria EQIP challenge.

\bibliographystyle{quantum}
\bibliography{biblio}

\end{document}